\documentclass[aps,pre,showpacs,showkeys,floatfix,unsortedaddress]{revtex4-1}
\usepackage{amssymb}
\usepackage{amsmath}
\usepackage{graphicx}
\usepackage{amsbsy}
\usepackage{color}
\usepackage{bm}
\usepackage[normalem]{ulem}

\begin{document}

\title{Phase diagram of heteronuclear Janus dumbbells}

\author{Patrick O'Toole}
\email{patrick.otoole@sydney.edu.au}
\affiliation{School of Chemistry, University of Sydney, NSW 2006, Australia}

\author{Achille Giacometti}
\email{achille.giacometti@unive.it}
\affiliation{Dipartimento di Scienze Molecolari e Nanosistemi, 
Universit\`{a} Ca' Foscari di Venezia,
Campus Scientifico, Edificio Alfa,
via Torino 155, 30170 Venezia Mestre, Italy}

\author{Toby Hudson}
\email{toby.hudson@sydney.edu.au}
\affiliation{School of Chemistry, University of Sydney, NSW 2006, Australia}

\date{\today}
\begin{abstract}
Using Aggregation-Volume-Bias Monte Carlo simulations along with Successive Umbrella
  Sampling and Histogram Re-weighting, we study the phase diagram of a system of dumbbells formed by two touching spheres having variable sizes, as well as different interaction properties. The first sphere ($h$) interacts with all other spheres belonging to different dumbbells
  with a hard-sphere potential. The second sphere ($s$) interacts via a square-well interaction with other $s$ spheres belonging to different dumbbells and with a hard-sphere potential with all remaining $h$ spheres. We focus on the region where the $s$ sphere
  is larger than the $h$ sphere, as measured by a parameter $1\le \alpha\le 2 $ controlling the relative size of the two spheres.
  As $\alpha \to 2$ a simple fluid of square-well spheres is recovered, whereas $\alpha \to 1$ corresponds to the Janus dumbbell limit, where the $h$ and $s$ spheres have equal sizes. Many phase diagrams falling into three classes are observed, depending on the
  value of $\alpha$. The $1.8 \le \alpha \le 2$ is dominated by a gas-liquid phase separation very similar to that of a pure square-well fluid with varied critical temperature and density. When $1.3 \le \alpha \le 1.8$ we find a progressive destabilization of the gas-liquid phase diagram by the onset of self-assembled structures, that eventually lead to a metastability of the gas-liquid transition below $\alpha=1.2$.
\end{abstract}

\maketitle

\section{Introduction}\label{introduction}
A class of particle, now by convention given the moniker `Janus', has attracted much attention in the last few decades.\cite{JIANG2012} Janus particles are characterised as possessing a patch which delineates regions of the particle with differing \emph{philos}. While spherical Janus particles have been studied extensively in experiments\cite{ROH2005,WALTHER2008,PAWAR2010,WALTHER2013} and theory,\cite{SCIORTINO2009,SCIORTINO2010} only recently dumbbell shaped Janus particles have attracted equal attention\cite{WHITELAM2010,MUNAO2014,AVVISATI2015} partially because they can now be synthetized rather routinely.\cite{KRAFT2012} 

In essence, Janus dumbbells can be reckoned as the analogue of molecular dimers at the colloidal scale, with the great advantage of having tunable interactions that are not constrained by stochiometry, so that the combined effect of maximizing the number of favourable contacts and optimizing the steric hindrance give rise to a rich polymorphism in their phase diagrams.
 
Different from Janus particles, having spherical symmetry in the particle shape and anisotropy in the interaction potential\cite{KERN2003}, Janus dumbbells are characterized by
a spherically symmetric potential and shape anisotropy.\cite{WHITELAM2010,MUNAO2014,AVVISATI2015} The simplest case of this class is given by homonuclear tangent beads, each characterized by a hard-sphere interaction complemented by a square-well attractive tail. This system demonstrates a conventional gas-liquid phase separation whose phase diagram depends essentially only on the interaction range of the square-well. As the attractive interaction on one site is gradually reduced, the coexistence region of the gas-liquid phase separation is progressively shrunk by the onset of micelles at low densities and temperatures, and lamellae at very low temperatures and higher densities.\cite{MUNAO2014} This trend persists until eventually gas-liquid phase separation becomes 
metastable to the formation of self-assembled structures upon approaching the `Janus limit', where the attractive interaction on one site is nil.\cite{MUNAO2014} This is the system that is usually referred to as Janus dumbbells. Work on similar systems has yielded structurally diverse self-assembly behaviour that depends upon the inter-nuclear distance and the interaction site diameter ratio.\cite{WHITELAM2010, AVVISATI2015} 

Janus dumbbells can then be generalized by allowing the two beads to have different sizes,\cite{MUNAO2015} the heteronuclear Janus dumbbells (HJD). This can be done in two ways, that give rise to rather different behaviours. As the size of the non-interacting bead (having only hard-sphere potential) is increased with respect to the attractive one (having the additional square-well tail), the phase diagram is dominated by the formation of self-assembled aggregates. This is a very interesting regime displaying a very rich and unconventional phase diagram that will be analysed in detail in a following article. The opposite case, where the size of the bead
that is the origin of the additional square-well attractive tail interaction is larger than the hard-sphere counterpart, is the focus of the present work. Under this condition, one expects a
competition between a gas-liquid phase separation, that is favoured when the square-well bead is much larger than the hard-sphere companion, and self-assembly behaviour that is conversely favoured when the two sizes are comparable and hence the Janus limit is approached. 

In a previous paper\cite{MUNAO2015} the full span of size asymmetry was scanned by computing the second-virial coefficient and the corresponding Boyle temperatures (the temperature at which attractive and repulsive interaction are comparable). Canonical ensemble Monte Carlo simulations \cite{FRENKEL2002} using conventional roto-translation moves were employed to underpin the interesting regimes in terms of size ratio between the (larger) square-well and (smaller) hard-sphere parts of the dumbbells, in terms of competition between phase separation and cluster formation.
In the present work, we focus on this regime by investigating firstly the liquid behaviour on approach to the symmetric case (previously referred to as the Janus limit) and secondly 
by establishing the approximate location and shape of self-assembled structures emerging at diameter ratios approaching the Janus limit. 
We employ Monte Carlo (MC) simulations utilising the Aggregation-Volume-Bias (AVBMC) particle move algorithm \cite{CHEN2000,CHEN2001} along with Successive Umbrella Sampling (SUS) to study the gas-liquid behaviour, and simulations in the canonical ensemble for structural characterisation. By a careful extrapolation of the gas-liquid critical point along the pathway leading to the symmetric case, we are able to identify the exact location of the putative critical point of the Janus limit \cite{MUNAO2014}.

This paper is structured as follows: In section \ref{model} we define the model employed for traversing the diameter ratio space; in section \ref{method} we describe the implementation of the SUS algorithm and the histogram re-weighting process, the Aggregation-Volume-Bias Monte Carlo algorithm and its application to the tangent Janus dumbbell with unequal site core diameters (HJDs); In section \ref{results} we discuss the gas-liquid critical phenomena with respect to the variation in site diameter ratio (\ref{res:liqgtone}), the structure of the corresponding liquid and characterisation at the onset of self-assembled bilayer structures (\ref{res:liqstruct}), discuss the structure of self-assembled phases observed (\ref{res:alpha_gt_one}), and present phase diagrams collating these data in \ref{res:phasediagrams}; finally we conclude in section \ref{conc}.

\section{Model parameters}\label{model}

Particle systems comprised of tangent HJDs with characteristic length parameter $\sigma$, composed of two spherical interaction sites, referred to as beads $ s $ and $ h $, with diameters, $ \sigma_s $ and $ \sigma_h $, each modified by a size-ratio parameter $\alpha \in [0,2]$. Parameter $\alpha $ defines the relative size of each bead core by

\begin{equation}
\begin{aligned}
&\begin{aligned}[c]
\sigma_s = \begin{cases}
\alpha\sigma &\text{$\alpha \leqslant 1$}\\
\ \sigma &\text{$\alpha \geqslant 1$}
\end{cases}
\end{aligned}
\\
&\begin{aligned}[c]
\sigma_h = \begin{cases}
\quad\sigma &\text{$\alpha \leqslant 1$}\\
(2-\alpha)\sigma &\text{$\alpha \geqslant 1$}
\end{cases}
\end{aligned}
\end{aligned}
\end{equation}

where $\sigma_h$ is the diameter of a purely repulsive hard-sphere (HS) and $\sigma_s$ is the diameter of the core of a square-well sphere (SW), with interaction range in addition to its hard-core, parametrized as $\lambda (= 0.5) $, such that the resultant interaction range is 

\begin{equation}
\gamma = \sigma_s+\lambda\sigma_s. 
\end{equation}

Neighbouring particles whose $ s $ sphere's centre comes within $ \gamma $ of the $ s $ sphere centre have a bonding energy of $-\varepsilon (<0)$. The total site-wise interaction potential between two particles is defined as

\begin{equation}
\begin{aligned}
V_{Total} =& V_{SW}(r_{s,s}) + V_{HS}(r_{h,s})\\
		   & + V_{HS}(r_{s,h}) + V_{HS}(r_{h,h}),
\end{aligned}
\end{equation}

\begin{figure}[htbp]
	\centering
	\includegraphics[width=15cm]{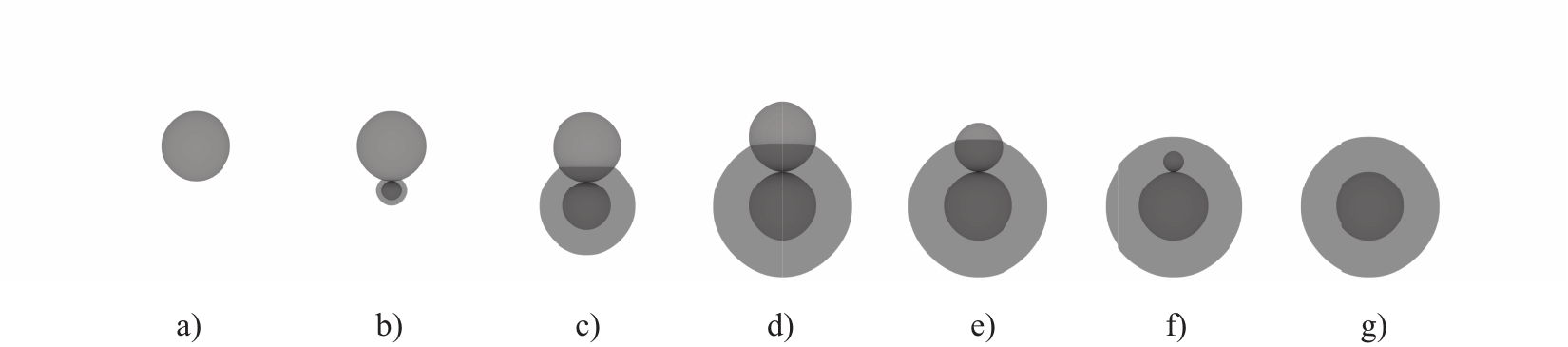}
	\par\bigskip
	\caption{Sketch of the particle at different points along the $ \alpha $ parameter: a) pure HS; b) $ \alpha = 0.25 $, such that $ \sigma_h = 1 $ and $ \sigma_s = 0.25 $; c) $ \alpha = 2/3 $; d) the Janus dumbbell where $ \alpha = 1.0 $; e) $ \alpha = 4/3 $, where $ \sigma_s = 1$ and $ \sigma_h = 2/3 $; f) $ \alpha = 7/4 $, where $ \sigma_s = 1 $ and $ \sigma_h = 1/4 $; g) pure SW. The shaded regions denote the range of the interaction of each $ s $ bead.}
	\label{variation}
\end{figure}

\emph{i.e.} as the sum of contributions from the $ s $ and $ h $ beads, where the potentials $ V_{SW} $ and $ V_{HS} $ are defined by

\begin{equation}
\begin{aligned}
& \begin{aligned}[c]
V_{SW}(r_{ss}) = \begin{cases}
\infty & \text{$ r_{ss} < \sigma_s $} \\
\!-\varepsilon & \text{$\sigma_s < r_{ss} \leqslant \gamma $}\\
\ 0 &\text{$r_{ss} > \gamma$}
\end{cases}
\end{aligned}
,\\
& \begin{aligned}[c]
V_{HS}(r_{ab}) = \begin{cases}
\infty & \text{$ r_{ab} < (\sigma_a + \sigma_b)/2 $} \\
\ 0 &\text{$r_{ab} > (\sigma_a + \sigma_b)/2 $}
\end{cases}
\end{aligned}
,\\
&\hspace*{10mm}\begin{centering}
(a,b) \in \{(s,h),(h,s),(h,h)\} ,
\end{centering}
\end{aligned}
\end{equation}\label{potential_definition}

where interaction sites along with the corresponding particle diameters. The particles therefore most simply, where $ \sigma_s = \sigma_h $, which we refer to as the Janus limit, take the form denoted in panel d) of Fig.\ref{variation}.

The interaction energy parameter $ \varepsilon $ is taken as the unit of energy and set to unity for all values of $ \alpha $. In this paper we study systems of particles where $1< \alpha <2 $,
the two limits of $\alpha=2$ and $\alpha=1$ corresponding to the square-well sphere and the homonuclear Janus dumbbell, respectively.
It is observed elsewhere\cite{MUNAO2015} that the region of $ 1 < \alpha < 2 $ can be roughly separated in two different regimes. In the first regime ($\alpha \geq 1.3$) the phase diagram closely resembles that of the square-well dumbbells\cite{VEGA1992} with scaled critical temperatures and densities, we here identify the exact $\alpha$ dependence of both. As the Janus limit is known to have a metastable critical point,\cite{MUNAO2014} it is fairly clear at some point along 
the $ \alpha $ pathway that gas-liquid phase separation may either coexist, or compete with the formation of bilayer structures as $ \alpha \rightarrow 1 $.\cite{MUNAO2015}
Here, we also  study the regime $1 < \alpha < 1.3$ in some detail and demonstrate that around $\alpha \approx 1.3$ a rather unconventional self-assembly process progressively takes place thus destabilizing 
gas-liquid phase separation. This new regime cannot be observed by conventional MC methods, and we have been employing a dedicated method to study it, that is described next.

\section{Methods}\label{method}

\subsection{Aggregation Volume Bias Monte Carlo}
A customized version of the Aggregation Volume Bias Monte Carlo (AVBMC) algorithm\cite{CHEN2000,CHEN2001} that facilitates an increased rate of not only bond breakage and formation but also facilitates cluster hopping was implemented, with the aim of improving the gathering of statistics for all densities and $\alpha$ studied as well as facilitating the rapid relaxation of liquids to equilibrium configurations in successive umbrella sampling (SUS) simulations that will be described next. The AVBMC algorithm involves choosing a particle from a bonded configuration (i.e. a particle whose centre is within the 'bonding volume' $V_{in}$ of a given particle) and translating it either to the non-bonding volume of that particle or the bonding volume of another particle, or taking a particle from outside the bonding volume of a particle and translating it to inside the bonding volume. The bonding volume used in this work is defined by

\begin{equation}
\begin{aligned}
V_{AVB} = & \frac{\pi}{6}\left[8\gamma^3-\sigma_s^3\right] \\
& \times \left(1-\left(\frac{1}{2}-\frac{\gamma}{2(\sigma_s+\sigma_h)}\right)\right).
\end{aligned}
\end{equation}

This parametrises a bonding region defined by the bond distance, $\gamma$, minus the volume of a spherical cone defined by the presence of the hard sphere component. 
Further details of the procedure can be obtained from the original paper \cite{CHEN2001}.

\begin{figure}[htbp]
	\includegraphics[width=10cm]{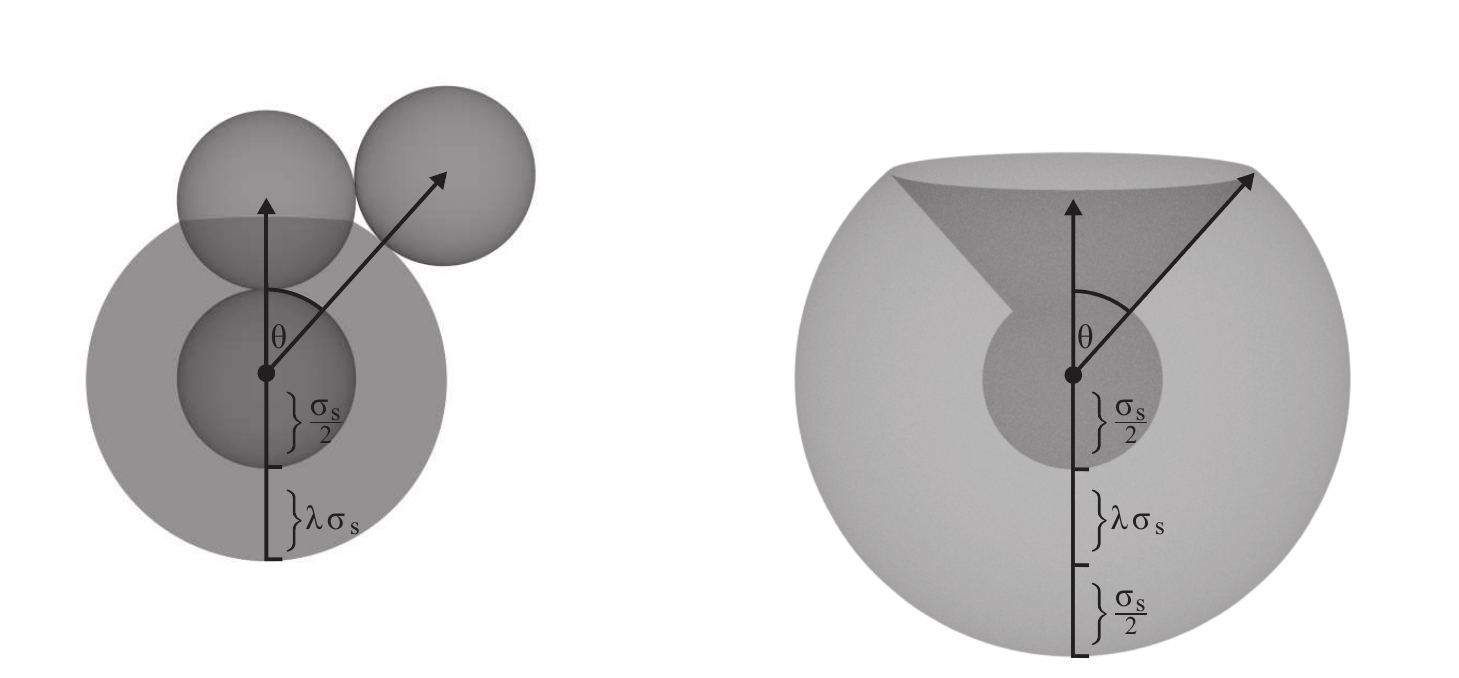}
	\caption{The \emph{in} region of a particle for use with AVBMC. It is defined by the furthest distance away an $ s $ bead can contribute to $ \mathcal{U} $. The volume consists of a layer of a sphere of radius $ \sigma_s+\lambda\sigma_s $ with a conical section subtracted as defined by angle from the nearest approach of a sticky sphere to the non-interacting sub-sphere.}
	\label{avbmcvolfig}
\end{figure}

\subsection{Successive Umbrella Sampling}

Successive Umbrella Sampling (SUS) extends a method for estimating free-energy\cite{TORRIE1977} whereby the range of states to be explored by the umbrella sampling procedure is restricted to windows of width $ \omega$, investigating windows one after the other such that the state space can be traversed without the need for a weight function, as is the case with multi-canonical approaches.\cite{VIRNAU2004} We employ Grand Canonical ensemble (GC) simulations of systems with each window  able to explore system particle numbers $ N $ and $ N+1 $ states. A histogram $ H_k[N] $ records the how often the simulation visits each state in the $ k^{th} $ window $ [k\omega,(k+1)\omega] $. The left and right bins of each histogram, $ H_{kr} = H _k[k\omega] $ and $ H_{kl} = H _k[(k+1)\omega] $, and their ratios $ r_k \equiv H_{kr}/H_{kl} $ can then be compiled to yield an un-normalised probability distribution

\begin{equation}\label{histogram_compilation}
\begin{aligned}
\frac{P[N]}{P[0]} &= \frac{H_{0r}}{H_{0l}} \cdot \frac{H_{1r}}{H_{1l}} \ldots \frac{H_k[N]}{H_{kl}} \\ 				  &=\prod\limits_{i=1}^{k-1}r_i \cdot \frac{H_k[N]}{H_{kl}}
\end{aligned}
\end{equation}

In the limit of small $ \omega $ individual simulations can be run in parallel (providing the computational resources are available), such that if the space is distributed over $ N_{proc} $ processors the resultant speed up for an MC algorithm with scaling $ \mathcal{O}(k\log k) $ will be proportional to $ (\sum_{k=1}^{N_{max}} k\log k) / N_{proc}$. 

\subsubsection{Histogram Re-weighing}\label{reweighist}
Histogram re-weighting is performed by segmenting the $ N $ space and the corresponding $ P(N) $ into regions corresponding to the different phases encountered over $ N $ and identifying minima in the $ P(N) $ which delineate regions of different phases, where $ P(N) $ is comparatively large. After locating satisfactory minima in $ P(N) $, the areas astride are compared to identify the direction which we must modify the distribution to yield equal areas (and thus equal volume of phase space). The process is carried out by adjusting the histogram using the chemical potential, $ \mu_{_f} $. Here we first switch our distribution to $ P(\rho) $ (which implies dividing each $ N_i $ by the simulation cell volume $ V $). The process is identical if unmodified from the $ N $ space, with the exception that the index in Eq.\ref{reweight_factor} is simply $ N_i $. At each $ \rho_i $, $ P(\rho_i) $ is modified by multiplying by a power of $ \mu_{_f} $ according to 

\begin{equation}\label{reweight_factor}
P'(\rho_i) = P(\rho_i)\mu_{\!{_f}}^{\rho_i},
\end{equation}

with total area normalisation, and the re-weighting process applied recursively until the compared regions are equal in area. The weight factor $ \mu_{\!{_f}} $ is modified by a single protocol, i.e. for coexisting phases $ \phi $ and $ \psi $, with areas in the $ P(\rho) $ distribution $ A_\phi $ and $ A_\psi $ and $ \rho_\phi > \rho_\psi $

\begin{equation}
\mu_{\!{_f}}' = \begin{cases}
\; (1 + \delta) \mu_{\!{_f}} & \text{$ A_\phi < A_\psi $}\\
\ (1 - \delta)\mu_{\!{_f}} &\text{$A_\phi > A_\psi $}
\end{cases}.
\end{equation}

A non-zero $ \mu_{\!_f} $ indicates that $ \mu \neq \mu_{coex}$.  If, after the re-weighting process, the factor by which we have modified the imposed $ \mu $ at the outset of the simulations lies outside the tolerance range of $ \mu_{\!_f} \in (0.98,1.02) $ the starting $ \mu $ is scaled by a factor proportional to $ \mu_{\!_f} $ and the simulation set begun anew. In practice, the compilation of the histogram can run into issues associated with memory underflow, where successive multiplications of low histogram ratios over regions of the density space with vanishing probability, for example intermediate densities between highly probable regions at low $ T^* $. To mitigate this problem, the compilation and re-weighting process can be performed in $ \log $ space with the operations altered appropriately.

\vspace{5mm}

\subsubsection{Implementation}
\begin{figure}[htbp]
	\centering
	\includegraphics[width=10cm]{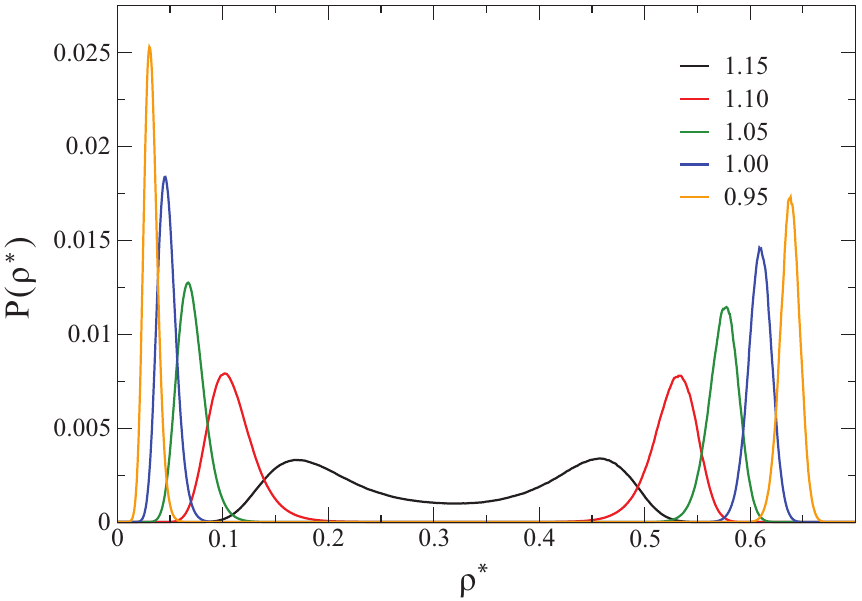} 
	\par\bigskip
	\caption{Binodal character developing for $ \alpha = 1.8 $ on lowering the temperature (indicated in the top-right corner) past the critical point ($ T^*_c \approx 1.156 $).}
	\label{Prvr1_8}
\end{figure}

We implement a version of the SUS protocol to study gas-liquid phase co-existence for $1<\alpha<2$. The SUS algorithm discretizes the density space into simulation windows ($\omega$) of width $N(=2$) such that each `edge' of the window overlaps with the adjacent window. In this way independent simulations are carried out utilising (GC) ensemble particle insertion and deletion moves. These histograms are compiled and, where necessary, re-weighted to yield points on the binodal via distributions of $ P(\rho^*) $ corresponding to the average density of the coexisting phases. Fig.\ref{Prvr1_8} displays the resultant $ P(\rho^*) $ distributions obtained via the SUS protocol, demonstrating the resolution of conventional gas-liquid phase separation on reducing the system temperature, $ T^* $, past the corresponding critical temperature, $ T^*_c(\alpha =1.8) \approx 1.156$. To obtain these distributions SUS simulations of particle systems up to and including $N=1000$ particles are equilibrated at constant volume. Standard periodic boundary conditions were employed throughout. We typically used between $10^6 $ and $10^7$ MC cycles for equilibration and additionally $10^6$ MC cycles in the production runs for collection of the statistics. A single constant volume MC cycle consists of $N$ trial single-particle moves combining a translation
of the dumbbell centre-of-mass and a rotation about a coordinate axis each drawn randomly with equal probability. Particle insertion and deletion moves in the GC have been carried out following standard prescription.\cite{FRENKEL2002} Equilibration runs are carried out 
to optimise the system energy, $ \langle U\rangle $, and system cluster statistics: the average number of monomers, $ \langle N_{mono}\rangle $; the average number of clusters, $ \langle N_c\rangle $; and the average cluster size, $ \langle N_s \rangle $; until fluctuations in each of these metrics was consistent with equilibrium, at which point GC insertion and deletion moves are employed to populate the histogram edges. Histogram edge ratios are monitored during each simulation to ensure convergence to a stable ratio, such that $ H(N+1)/H(N) \approx k \pm 0.001$ (where $ k $ is the converged histogram ratio), before the histogram is compiled utilising Eq.\ref{histogram_compilation}. Once the histograms are compiled a re-weighting technique is applied and the resultant densities of the co-existing gas and liquid, and their relative errors obtained. 

\subsubsection{Critical points}

Critical values $ \rho^*_c $ and $ T^*_c $, the system number density and temperature at the critical point, are obtained by fit using a formulation of the law of rectilinear diameters \cite{FRENKEL2002},

\begin{equation}\label{rectilin_a}
\frac{\rho^*_{_l} + \rho^*_{_g}}{2} = \rho^*_c + A(T^*-T^*_c)
\end{equation}

where $ \rho^*_{_l} $ and $ \rho^*_{_g} $ are the average number density of the liquid and gas phases at coexistence, $ T^* $ is the system temperature, and $ A $ is a fitting parameter. The density difference at coexistence $ \rho^*_{_l} - \rho^*_{_g} = \Delta_{l-g} $ is fit using a scaling law,

\begin{equation}\label{rectilin_b}
\Delta_{l-g} = B(T^*-T^*_c)^{\beta_c},
\end{equation}

where $ B $ is a fitting parameter and $ \beta_c $, the critical exponent, fixed to the Ising class $ \beta_c = 0.32 $. Tab.\ref{crit_table} summarises the parameters obtained from the fit. Fig.\ref{coex_by_alpha} shows the variation in the density co-existence curves for $ 1.2 \leqslant \alpha \leqslant 2.0 $. 

\section{Results}\label{results}

\subsection{Boyle temperature}
\begin{figure}[htbp]
	\centering
	\includegraphics[width=10cm]{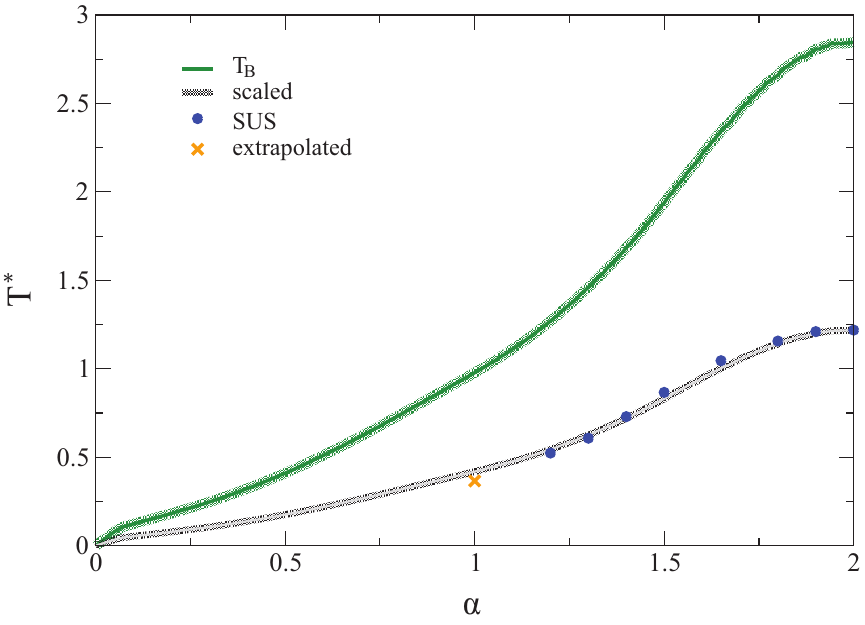} 

	\par\bigskip
	\caption{Variation in $ T_B $ with respect to $ \alpha $. The green line indicates the Boyle temperature, $ T_{\mathcal{B}} $. with the maximum error indicated by the corona. The grey line indicates the variation in $ T_B $ when scaled to meet the calculated SW critical point; blue dots are $ T^*_c(\alpha) $ calculated by SUS in this work.}
	\label{Boyle_T_valpha}
\end{figure}

In order to guide the exploration of critical phenomena over the $ \alpha $ parameter we use the result of a numerical calculation \cite{MUNAO2015} of the Boyle temperature, $ T_{\mathcal{B}} $, the temperature at which the second coefficient of the virial expansion changes sign. Scaling $ T_{\mathcal{B}} $ to the value of the critical temperature of a pure SW fluid (\emph{I.e. $ \alpha = 2 $}) is a simple heuristic to estimate the variation in $ T_c^*(\alpha) $. Beginning where $ \alpha = 2 $, the SUS protocol is applied to track variations in critical phenomena. Fig.\ref{Boyle_T_valpha} demonstrates the quality of predictive power of $ T_{\mathcal{B}} $ to the critical temperature of the dumbbells at various $ \alpha $ observed.

\vspace{5mm}

\subsection{Gas-liquid Phase Co-existence}\label{res:glphasecoex}
\begin{figure}[htbp]
	\centering
	\includegraphics[width=10cm]{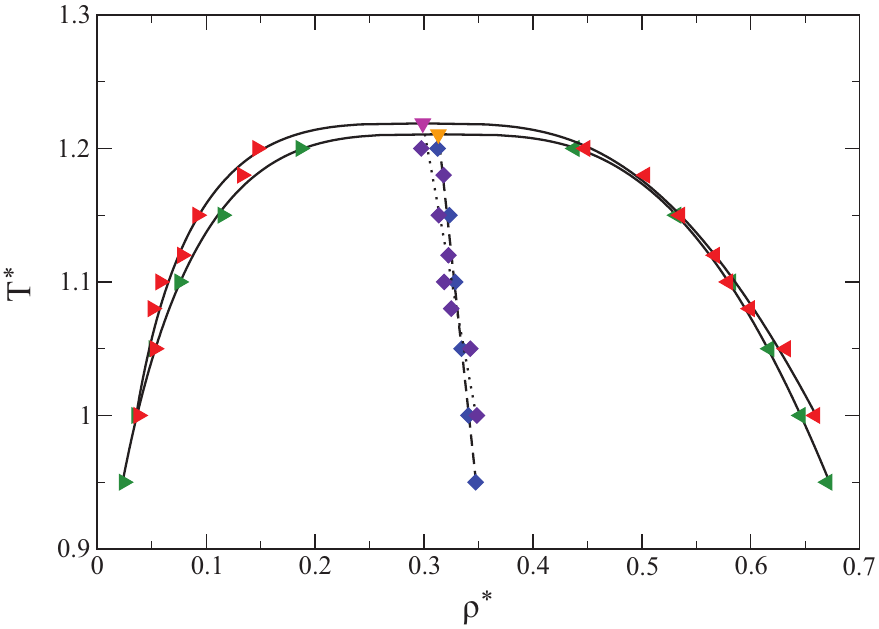} 
	\par\bigskip
	\caption{Comparison between the gas-liquid coexistence curve as calculated by GEMC (red and indigo symbols with magenta critical point) and the SUS method (green and blue symbols with orange critical point). Reasonable agreement between the two estimates is demonstrated, however a small differences in $ \rho^*_c $ and $ T^*_c $ are observable. More weight can be applied to the SUS technique since it is less prone to systematic error in the density near the critical point.}
	\label{coex_vega_comp}
\end{figure}

In order to test our algorithm, we start from the square-well dumbbell case, \textit{i.e.} $ \alpha = 2 $ where there are only $s$ spheres , with $ \lambda = 0.5 $ where the gas-liquid phase diagram was obtained using Gibbs ensemble Monte Carlo (GEMC) simulations.\cite{VEGA1992} The value $\lambda=0.5$ will be kept fixed in all simulations since it can be reckoned as a reasonable compromise between very short-range fluids 
($\lambda \lesssim 0.15$) , requiring
very demanding and extensive numerical simulations in order to cope with the extremely low temperatures involved, and mean-field-like fluids $(\lambda \approx 1)$, that belong to a different universality class.
This value was also used in past work both on Janus particles\cite{SCIORTINO2009,GIACOMETTI2009,SCIORTINO2010,GIACOMETTI2010,GIACOMETTI2014} and Janus dumbbells\cite{MUNAO2013,MUNAO2014,MUNAO2015}, although
it should be stressed that this value is significantly larger than that found under typical experimental condition\cite{WALTHER2008,PAWAR2010,KRAFT2012,HU2012,WALTHER2013} that is of the order of $\lambda \approx 0.1$.
Fig.\ref{coex_vega_comp} compares coexistence curves obtained both under the current method and that obtained by GEMC\cite{VEGA1992} (with data refit according to Eq.\ref{rectilin_a} and Eq.\ref{rectilin_b}). Reasonably good agreement between the two methods is observed, with less than 1\% difference between estimates of $ T^*_c $. The deviation in $ \rho^*_c $ is significantly larger, closer to 5\%, but within error of the GEMC. The origin of this deviation is not clear but it is likely to be attributed to the large error bars in the GEMC.\cite{VEGA1992} We speculate that as density fluctuations become larger using the GEMC technique on approach to the critical point, estimation of the coexistence densities may lead to comparatively less reliable data than the SUS technique, which can simulate almost up the critical point providing the system is sufficiently large to capture enough of the diverging correlation length. 

\begin{figure}[htbp]
	\centering
	\includegraphics[width=10cm]{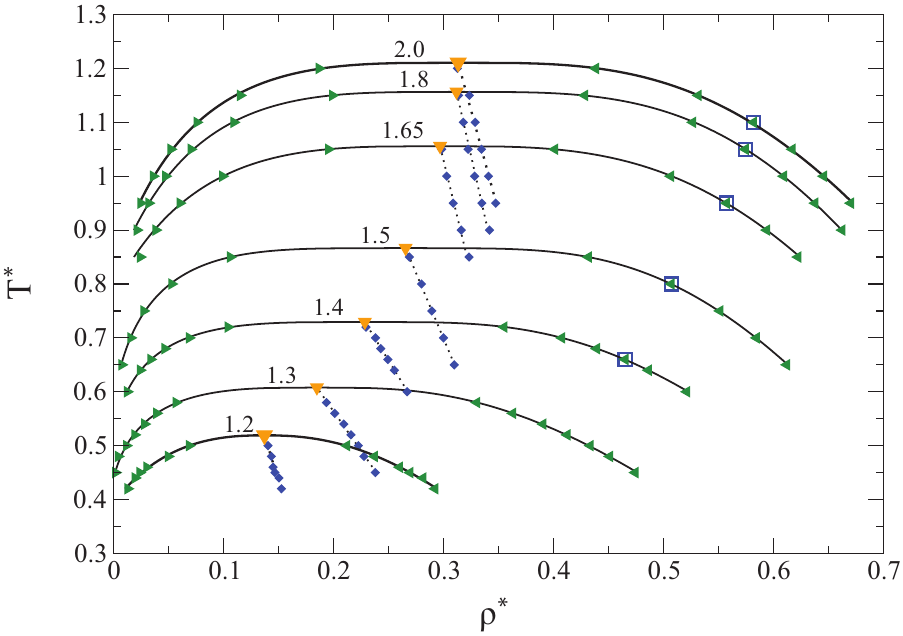}
	\par\bigskip
	\caption{Variation in the gas-liquid coexistence curves at different values of $ \alpha $ (indicated at the top of each fit). The critical points are indicated by the orange triangles, the green rightward and leftward pointing triangles indicate the densities of a coexisting gas and liquid, respectively. The blue squares indicate temperatures and densities at which the $ \mathbf{g}(r) $ is computed, chosen such that $ \phi^* $ is approximately equal to mitigate density effects, and displayed in Fig.\ref{gofr_byalpha}.}
	\label{coex_by_alpha}
\end{figure}

Moving away from the spherically symmetric case of $ \alpha = 2 $ is effectively that of swelling the $ h $ bead on the surface of the core of the $ s $ bead (see Fig. \ref{variation}). Fig.\ref{coex_by_alpha} demonstrates the outcome of SUS simulations at selected temperatures over the density space for each $ \alpha $. 

\begin{figure}[htbp]
	\centering
	\includegraphics[width=15cm]{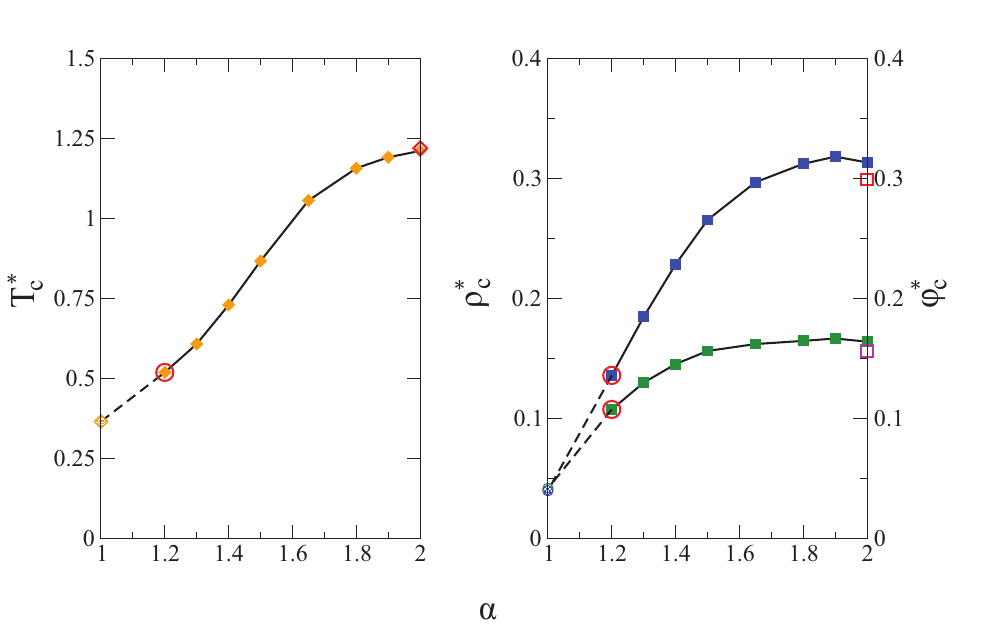} 
	\par\bigskip
	\caption{$ T^*_c $, $ \rho^*_c $, and $ \phi^*_c $ against $ \alpha $. The differently shaded symbol at $ \alpha = 1 $ indicates the location of the projected critical point from the study documented in Ref. \cite{MUNAO2014}. The dotted line indicates a linear interpolation between the last point at which a liquid is observed by SUS ($ \alpha = 1.2 $), and the projected critical point. The green symbols indicate $ \phi^*_c $, with the corresponding magenta square the value calculated from Ref. \cite{VEGA1992}.}
	\label{criticals_by_alpha}
\end{figure}

\vspace{5mm}
The critical temperature, beginning at $ T^*_c = 1.211 (\pm 0.002) $ for $ \alpha = 2 $, in reasonable agreement with \cite{VEGA1992}. Exploring the effect of a swelling $ h $ bead, which for values of $ \alpha > 1.5 $ can fit inside the bonding volume of the $ s $ bead, one can observe little difference in the shape of the coexistence curve, with a monotonic decrease of the critical temperature for $1.2 \lesssim \alpha \lesssim 2 $. This feature is typical of Janus systems\cite{SCIORTINO2010,GIACOMETTI2014}, and can be easily rationalized in terms of the diminishing volume of the interaction range that will limit the coordination number and thus decrease the temperature at which a critical point may be observed. On the other hand, $ \rho^*_c $ shows a small increase with respect to the pure phase where $ 1.8 \leqslant \alpha < 2.0 $. This behaviour is discussed further below. On decreasing $ \alpha $ below $ 1.65 $ the coexistence curves lose their symmetry (consider the shapes of the curves in Fig.\ref{coex_by_alpha}). While both branches between $ 1.2 < \alpha < 1.5 $ shift to significantly lower density, the gas branch does so at a faster rate as can be seen by the increasing slope of $ (\rho^*_l+\rho^*_g)/2 $. By $ \alpha = 1.5 $, the gas branch has shot off to a far lower density, in a trend that continues until $ \alpha = 1.3 $, dragging with it the critical point. For $ \alpha \rightarrow 1.2 $, the decline in $ \rho^*_c $ is approximately linear, whereas over the full range of $ \alpha $ studied for critical phenomena, the variation in $ T^*_c $ appears sigmoidal, consistent with the variation in $ T_{\mathcal{B}} $. The green points on the right-most panel of Fig.\ref{criticals_by_alpha} represents the critical volume fraction $ \phi^*_c $. This is obtained by multiplying $ \rho^*_c $ by the volume of the dumbbell via equation 

\begin{equation}
\phi^*_c = \frac{\pi}{6}\left[\sigma_s^3+\sigma_h^3\right]\rho^*_c.
\end{equation}

As is evident from Fig. \ref{criticals_by_alpha}, the volume fraction possesses a very slight positive slope over the region $1.5 < \alpha < 2$, excepting the small nodule between $1.8 < \alpha < 2.0$. Where $ \alpha < 1.5 $, $ \phi^*_c $ decreases more rapidly, until the progression comes to the end of the critical parameters curve as calculated. 

Interestingly, the location of the projected critical point obtained by extrapolating values of Fig. \ref{criticals_by_alpha} is in agreement with that obtained in past work
\cite{MUNAO2014} via a different extrapolation pathway, thus strongly suggesting the reliability of this value as a putative critical point of the Janus dumbbells. 

\begin{table*}[htbp]
\small
	\caption{Summary of critical point fitting parameters obtained from non-linear fitting of the SUS coexistence data.}
	\label{crit_table}
  \begin{tabular*}{\textwidth}{@{\extracolsep{\fill}}llllll}
    \multicolumn{6}{c}{\textbf{Phase Separation Data}} \\
    	\hline
    	\multicolumn{4}{c}{Critical Parameters} & \multicolumn{2}{c}{Fitting Parameters} \\
    	\hline
    	$ \alpha $ 	& $ T^*_c $ & $ \rho^*_c $ 	& $ \phi^*_c $	& $ A $ 	& $ B $ 	\\
    	\hline
    	2.00   			& 1.210	 (0.004)	& 0.313	 (0.002)	& 0.163	(0.001)	& -0.1328	& 0.9671	\\
    	1.90 			& 1.180  (0.005)	& 0.318  (0.003)	& 0.166	(0.002)	& -0.1103   & 0.9343	\\
    	1.80 			& 1.156	 (0.005)	& 0.312	 (0.002)	& 0.164	(0.001)	& -0.1109 	& 0.9450 	\\
    	1.65 			& 1.055	 (0.006)	& 0.296	 (0.002)	& 0.162	(0.002)	& -0.1189	& 0.9771	\\
    	1.50	 		& 0.866	 (0.007)	& 0.265	 (0.003)	& 0.156	(0.003)	& -0.2055	& 0.8788	\\
    	1.40			& 0.729	 (0.004)	& 0.228	 (0.001)	& 0.145	(0.001)	& -0.3008	& 0.8932	\\
    	1.30  			& 0.607	 (0.007)	& 0.184	 (0.006)	& 0.129	(0.004)	& -0.3416	& 0.8579	\\
    	1.20  			& 0.519	 (0.011) 	& 0.131	 (0.008)	& 0.107	(0.007)	& -0.1518	& 0.7466	\\
    	\hline
  \end{tabular*}
\end{table*}

\begin{figure}[htbp]
	\centering
	\includegraphics[width=15cm]{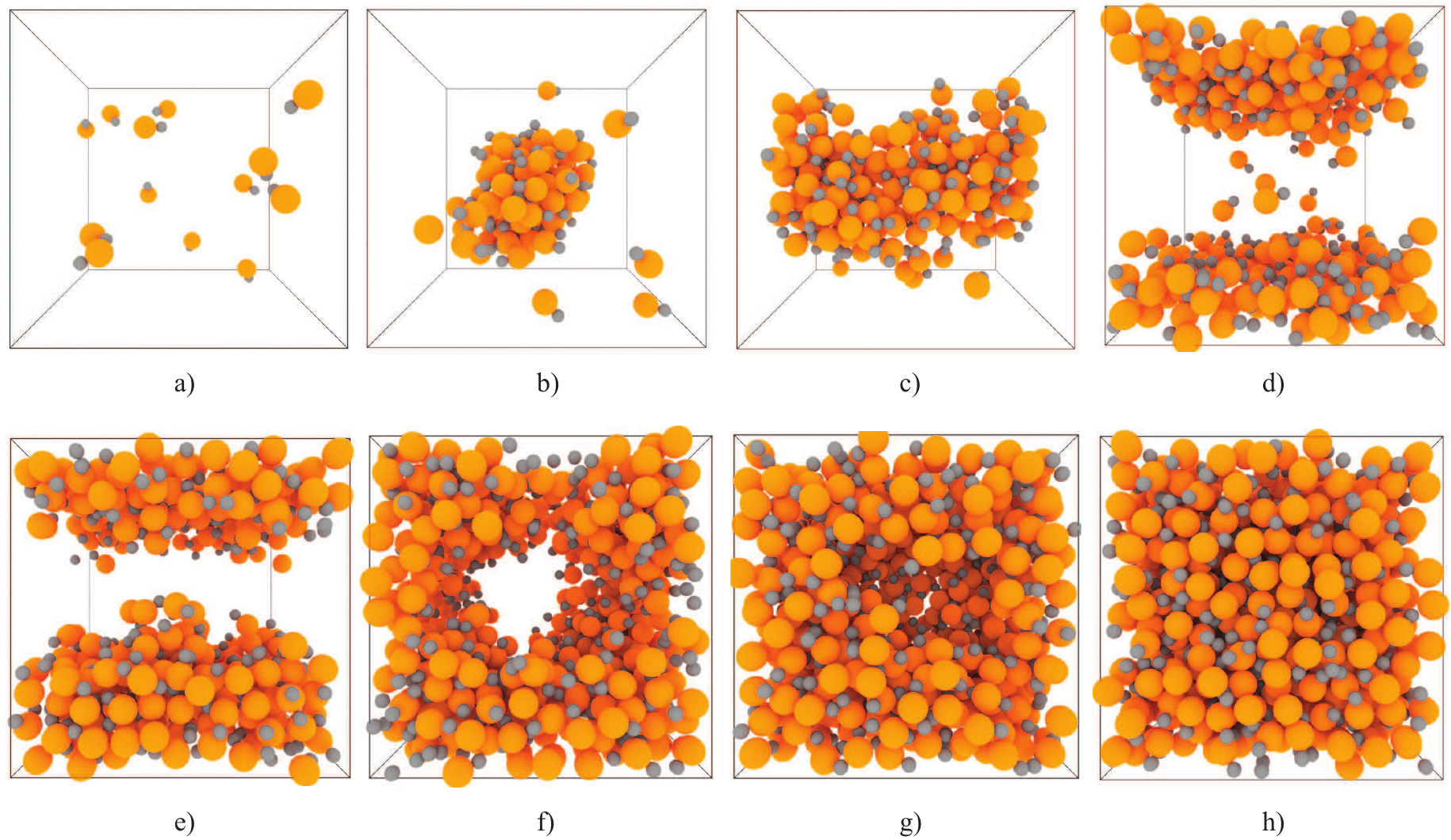} 
	\par\bigskip
	\caption{Snapshots of selected configurations obtained via AVBMC and SUS where $ \alpha = 1.5 $, $ T^* = 0.65 $. Here the SW-$ s $ beads are coloured orange, and the HS-$ h $ beads are coloured grey. These snapshots demonstrate the effects of the finite size of the simulation box usually observed in SUS simulations. They include: a) monomer gas $ \rho^* \approx 0.01 $; b) droplet coexisting with gas at $ \rho^* \approx 0.07$; c) percolated cylinder coexisting with gas at $ \rho^* \approx 0.13$; d) slab coexisting with gas $ \rho^* \approx 0.23$; e) slab at $ \rho^* \approx 0.31$; f) cylindrical bubble at $ \rho^* \approx 0.41$; bubble cavity at $ \rho^* \approx 0.5$; homogeneous liquid at $ \rho^* \approx 0.61$. }
	\label{1_5_liq_snaps}
\end{figure}

\subsection{Liquids of Janus Dumbbells}\label{res:liqstruct}

The small increase in $ \rho^*_c $ with $ 1.8 < \alpha < 2.0 $ is an unexpected result and warrants some analysis. One may consider, via a simple mean field style argument, that the presence of the $ h $ component ought to be interpreted as a reduction in the volume of the potential available for bonding. Adopting this view would lead one to infer a slight decrease in the temperature required to condense a liquid, and that this may be accompanied by an increase in $ \rho^*_c $, like the pure SW liquid on decreasing $ \lambda $.\cite{VEGA1992} However, the correction to the density anomaly by $ \alpha \approx 1.8 $ seems to indicate more than a single contributor to this density variation. To investigate the influence of the growing $ h $ bead, simulations of 1000 particles are performed in the canonical ensemble on systems at $ T^*<T^*_c $ and liquid $ <\!\rho^*_{coex}\!> $ across the range $1.4 < \alpha <2 $ to characterise any microscopic variation (the particular state-points examined are highlighted by the blue squares in Fig.\ref{coex_by_alpha}, chosen such that their $ \phi^* $ are approximately equal to mitigate the effects of density on the radial distribution function). Figure \ref{1_5_liq_snaps} displays some representative snapshots obtained for $\alpha=1.5$ at reduced temperature $T^{*}=0.65<T_c^{*}=0.85$ below the critical temperature and at increasing densities. Simulations of $ < 10^6 $ Monte Carlo sweeps (MCS) were sufficient to equilibrate these systems. Production sampling of the site-wise $ \mathbf{g(r)} $ is then performed over $ 2 \times 10^6 $ MCS.

\subsubsection{Structural Changes $ 1.4 \lesssim \alpha < 2.0 $}\label{res:liqgtone}

\begin{figure}[htb]
	\centering
	\includegraphics[width=10cm]{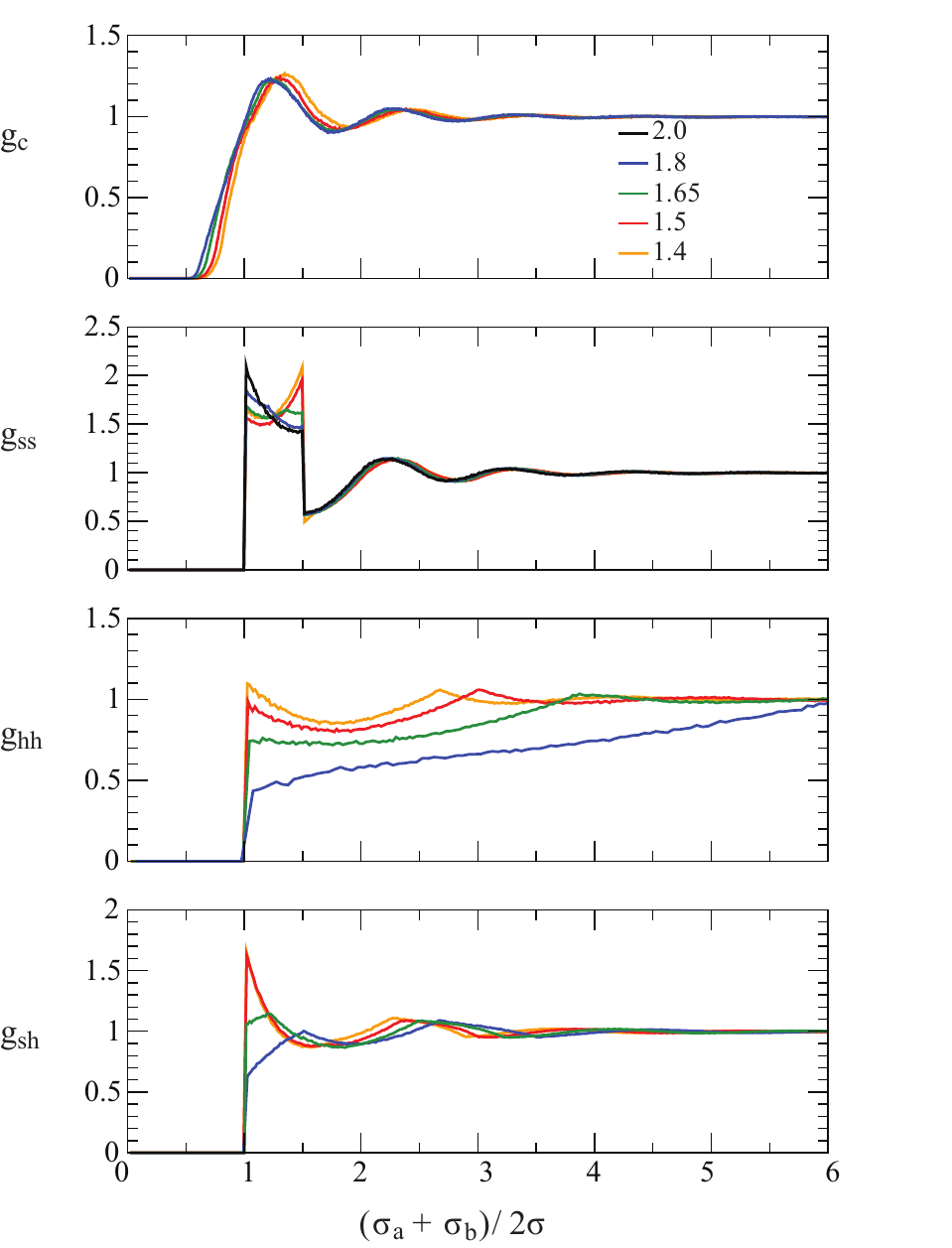} 
	\par\bigskip
	\caption{Distributions of site-wise $\mathbf{g ( r )}_{ab} $, where $ a,b \in {s,h} $, of liquids at coexistence densities formed at sub-critical temperatures over the range $ 1.4 < \alpha < 2.0 $. Colours indicate the value of $ \alpha $ indicated in the legend in panel 1 (note: $ \alpha = 2.0 $ is only present in panel 2).  These distributions demonstrate the effect of the presence of the $ h $ bead on the microscopic structure of the liquid. Further discussion of features can be found in the text.}
	\label{gofr_byalpha}
\end{figure}

The pair correlation functions in Fig.\ref{gofr_byalpha} characterise the average microscopic structure around each particle. The top panel, the centroid correlation $\mathbf{g}_{c} $, shows a slight elongation of the mode of all peaks for $ 1.4 < \alpha < 2$, indicating that as $ \sigma_h $ increases, the average $ c-c $ inter-particle distance increases. The second panel, the $ s $ sphere $\mathbf{g}_{ss} $, shows a gradual progression of the average position of $ s $ beads from the inner extent of the interaction range to the outer extent. One can observe for distances between $ \sigma_s $ and $ \sigma_s + \lambda\sigma_s $ the presence of, at first, a sharp peak at $ \sigma_s $ for $ \alpha = 2 $, which decays  until $ \sigma_s + \lambda\sigma_s $, where it drops significantly. The converse is true for $ \alpha = 1.5 $ and $ 1.4 $, where the opposite progression occurs. At intermediate $ \alpha $, the presence of $ h $ beads perturbs the average bonding environment around the $ s $ beads, leading to an additional peak or shoulder observable at $ \sigma_s+\sigma_h $ for $ \alpha \in \{1.8,1.65\} $. At the same time one can also observe the increasing correlation of the $ h $ components in the third panel and the $ s $ and $ h $ components in the fourth panel of Fig.\ref{gofr_byalpha} where the correlation of the $ s $ and $ h $ spheres increases. These data indicate that the $ h $ beads begin to play a significant role in the local environment around each bonding site at any $ \alpha $ away from $ 2 $, and that they begin to \emph{push} against neighbouring $ s $ beads, eventually reducing the number of bonds each particle makes. While the mean field interpretation gives us some insight as to why $ \rho^*_c $ increases slightly on a small increase of $ \sigma_h $, eventually the increasing correlation of the $ h $ beads begins to significantly perturb the local structure, leading to shifts in the distributions of particle positions in the bonding region in turn causing $ \rho^*_c $ to shift back toward the large $ \sigma_h $ behaviour. The growth of $ \sigma_h $ on decreasing $ \alpha $ eventually restricts the number of bonds per particle and increases the average bond length, such that in order to condense a liquid the system must be cooler. This kind of structural behaviour also seems to occur with structurally related particle types which are observed to present this inner-outer bond distance exchange with increasing particle anisotropy \cite{AVVISATI2015}. 

\subsection{Bonding Networks and Interfaces}\label{res:liqnetwork}

Where the critical parameters begin to drop rapidly for $ 1.5 < \alpha < 1.65 $ (Fig.\ref{criticals_by_alpha}), bonds formed across the $ h $ bead diameter are restricted to the outermost extent of the potential range, significantly altering the co-ordination of bonds around each $ s $ site. This leads, at sufficiently low temperature, to the formation of $ h $ rich pockets in the liquid since maximising the number of bonding interactions creates a drive to segregate the $ h $ components. As $ \sigma_h $ grows beyond $ \lambda\sigma_s $, the presence of $ h $ rich pockets grows, until the formation of bilayer structures occurs. Bilayer structures form where the presence of the $ h $ bead occupies enough of the bonding region to force a significant proportion of the $ h $ beads into the interface. 

\begin{figure}[htbp]
	\centering
	\par\smallskip
	\includegraphics[width=18cm]{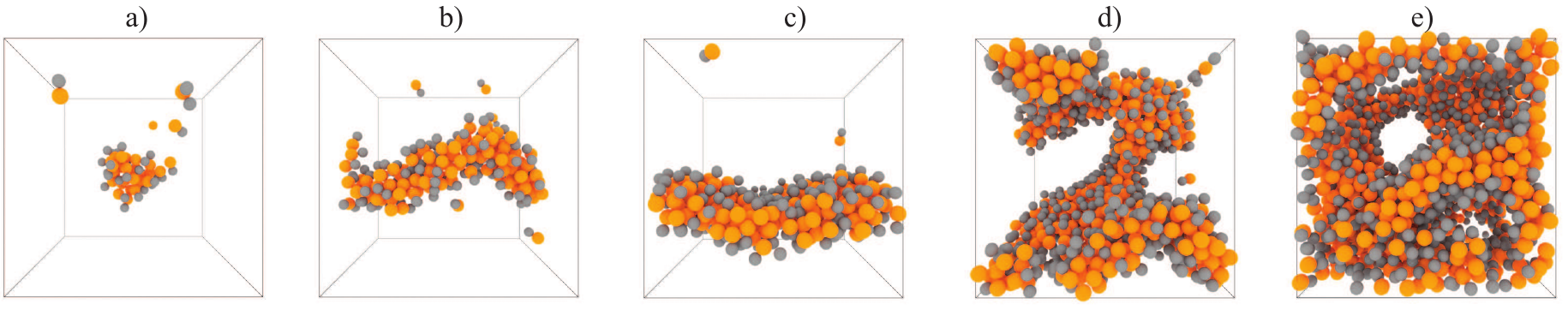} 
	\caption{Structures observed in SUS runs where $ \alpha = 1.2 $ with $ \omega_{max}=2000 $ across $ \rho^* $ at $ T^* = 0.42 $. From left: micelles at $ \rho^* \approx 0.01 $ (a); percolated string at $ \rho^* \approx 0.05 $ (b); percolated bilayer at $ \rho^* \approx 0.13 $ (c); curved bilayer slab with bridging arm at $ \rho^* \approx 0.25 $ (d); and a continuous cavity (percolated void) in the bilayer network liquid at $ \rho^* \approx 0.3 $ (e).}
	\label{finite_size_snaps}
\end{figure}

\begin{figure}[htbp]
	\centering
	\includegraphics[width=12cm]{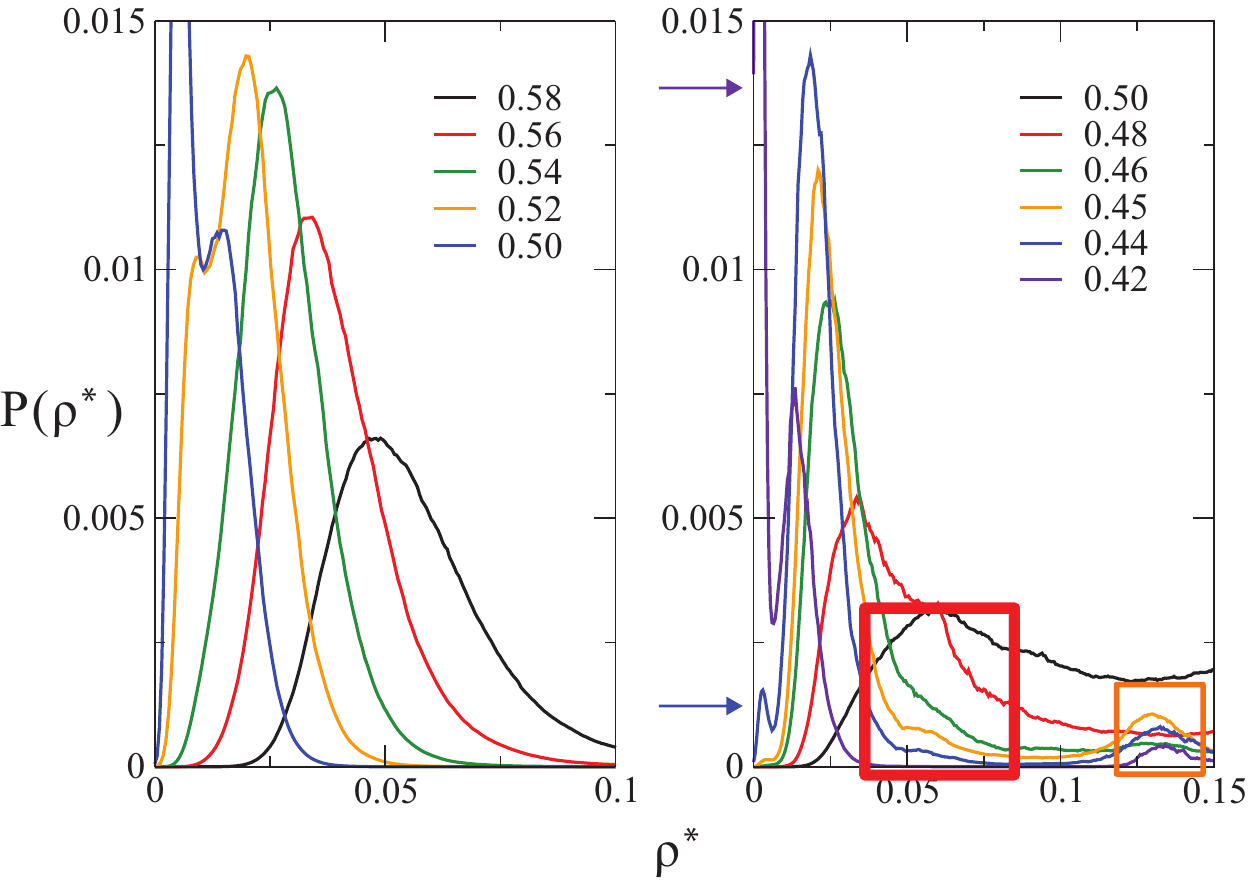} 
	\par\bigskip
	\caption{Finite size effects on the coexisting gas branches of the $ P(\rho^*) $ against $ \rho^* $ for $ \alpha = 1.3 $ (left) and $ \alpha = 1.2 $ (right). Additional peaks and shoulders manifest over certain density ranges due to the finite size of the simulation box and the peculiarities of the potential description. Coloured boxes indicate the locations in the $ \rho^* $ space of finite size. The orange box captures a region where a single bilayer has percolated across the cell diameter (see panel c of \ref{finite_size_snaps}); the red box indicates a region where a cylinder percolates across the cell(see panel b of \ref{finite_size_snaps}); the blue arrow indicates the monomer gas peak which is metastable with respect to the small nucleate peak where $ \rho^* \approx 0.025 $ at $ T^* = 0.44 $, but dominates at $ T^*= 0.42 $ (violet arrow).}
	\label{finite_sizePn}
\end{figure}

This behaviour causes two problems for simulation. Firstly, at high liquid densities the propensity of the particles to align such that their $ s $ beads face inward from an interface and their $ h $ beads face outward toward the interface by any layered structure implies that the number of insertion sites with $ -\Delta U $ at low temperature, where the interface has adopted a concave structure --- such as with the aforementioned $ h $ rich regions is depleted, rendering the acceptance of insertion moves low. This is clearly illustrated by Figure \ref{finite_size_snaps} reporting representative snapshots of configurations at $\alpha=1.2$, $T^{*}=0.42<T_c^{*}=0.512$ and increasing densities. Particularly noteworthy appears panel e) showing a continuous cavity in a a bilayer network liquid. Secondly, any nucleated structure will be affected by the finite size of the simulation box as particles tend to form interfaces with their hard $ h $ bead facing the void. Formation of elongated structures cause percolation to occur at low density, quite close to the gas peak. Fig.\ref{finite_sizePn} demonstrates the effect of finite size on the gas branch of the $ P(\rho^*) $ curves for $ \alpha \in {1.3,1.2}$. One may consider that these structures are thermodynamic minima in the density space, although careful inspection of the state-points must be performed to ascertain their properties. It is demonstrated elsewhere that for simulations performed to compute the coexistence densities via MC techniques in the GC ensemble that the finite size of the simulation cell stabilises structures that minimise their interfacial free energy \cite{BINDER2012}. I.e. on the scale of a finite simulation, intermediate phases which possess minimal surface area are thermodynamically stable, but may not be representative of the bulk behaviour in the thermodynamic limit (as $ N \rightarrow \infty $ and $ \phi^*_{interface} << \phi^*_{bulk} $). The remedy is to increase the size of the simulation sufficient to remove the influence on the binodal of the locally stable structures with respect to the coexisting gas or liquid.

\begin{figure}[htbp]
	\centering
	\includegraphics[width=15cm]{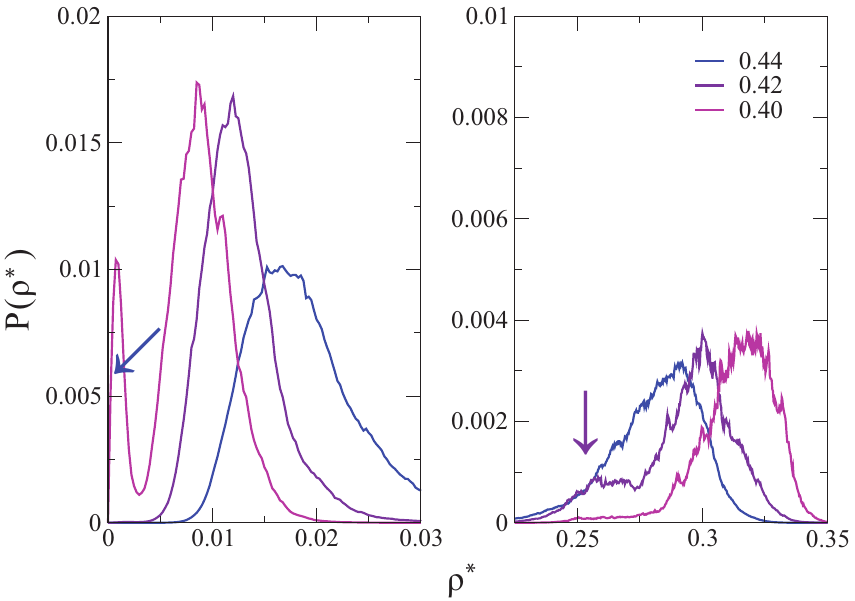} 
	\par\bigskip
	\caption{Persistence of finite size effects in simulations with larger system sizes in the binodal region for $ \alpha = 1.2 $. $T^*$ here indicated in the top of the right-hand panel. Arrows here indicate the position of finite size effects in the simulations. }
	\label{finite_sizePn_large}
\end{figure}

In the case of $ \alpha = 1.3 $, a simple system size increase is sufficient to remove the influence of the low density structures. For these systems, using a maximum window of $\omega=2000 $ and the corresponding box length such that $ \rho^*_{max} = 0.6$ is met by the final window (an effective $ N $-scale doubling). For $ \alpha = 1.2 $, the case is not so simple. Doubling the system size yields additional finite size effects (as can be observed in Fig.\ref{finite_sizePn_large}), causing problems for the sampling of the histogram bin edges. 

\subsection{Self-Assembled Structures}\label{res:alpha_gt_one}
\begin{figure}[htbp]
	\centering
	\includegraphics[width=15cm]{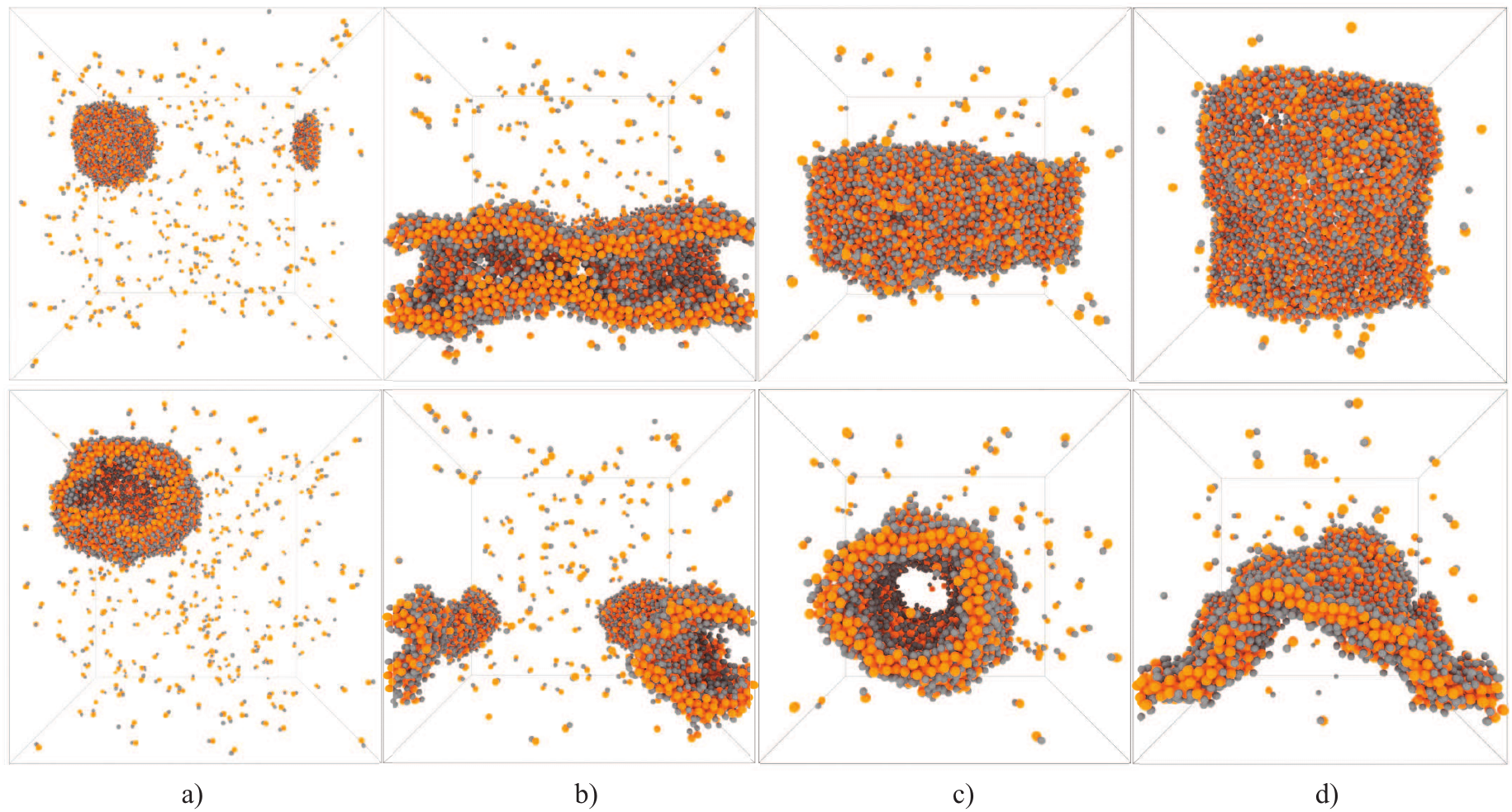} 
	\par\bigskip
	\caption{Structures obtained utilising the AVBMC algorithm at constant volume at $ \alpha = 1.2 $. The top and bottom each correspond to different aspects of the same snapshot: a) hollow vesicle coexisting with a monomer gas at $ \rho^* = 0.01 $; b) a percolated tube coexisting with a monomer gas at $ \rho^* = 0.03 $; c) tube with a larger diameter at $ \rho^* = 0.06$; a continuous wavy lamellar sheet at $ \rho^*= 0.07 $.}
	\label{large_sys_snaps}
\end{figure}

On approaching the Janus limit ($\alpha=1$), phase separation becomes progressively destabilized as indicated by Figure \ref{criticals_by_alpha}. In the lowest
asymmetric value considered here ($\alpha=1.2$),
the presence of highly structured percolated structures at very low density (Fig.\ref{finite_size_snaps}) implies that the characteristic length-scale is larger than the box dimensions, i.e. $ \sigma_{\epsilon} > V^{1/3} $. To explore whether $ \sigma_{\epsilon} $ is divergent or simply larger than the current $ L_{box} $, systems of $ N=3000 $ particles were simulated at constant volume employing the AVBMC algorithm at $ T^* \in {0.42, 0.44, 0.46} $ over the density range $ {0.01, 0.02, 0.03, 0.04, 0.05, 0.06, 0.07, 0.08} $. Snapshots of the self assembled structures can be seen in Fig.\ref{large_sys_snaps}.

At $0.01 \leqslant\rho^* \leqslant 0.03$, we observe the presence of a single aggregate structure, a vesicle coexisting with a monomer gas. Upon increasing the system density this vesicle structure percolates in 1D across the periodic boundary forming a tube (where $ 0.03 \leqslant \rho^* < 0.06) $), whose diameter increases with further increasing density to eventually percolate in a second dimension to form a wave-bilayer structure (where $ \rho^* \geqslant 0.07 $), the structure at $ \rho^* = 0.08 $ (see panel (d) of Fig.\ref{large_sys_snaps}). If one were to perform a constant pressure simulation across this isotherm, it may be the case that the vesicle and tube structures would disappear (since there is no barrier to surface merging imposed by the presence of the smaller $ h $ particle and one would obtain solely continuous layered structures. This is in contrast to vesicle structures observed elsewhere,\cite{SCIORTINO2009, AVVISATI2015} where the hard-core bead forms what is essentially a non-interacting shell around the vesicle. This possibility was not explored here. The observation of bilayer vesicles and a continuous tube with hollow internal cavities and curved sheet structures (since the smaller $ h $ bead allows the layer to tolerate some curvature) at such low densities is an important finding that may be of technological interest. While properly implemented PBC should return the behaviour of the bulk, the cubic cell geometry still exerts an influence on the characteristic length of any assembled structure, it is therefore the case that systems obtained in this region of the $ \alpha $ space --- i.e. where continuous structured systems occur at low $ \rho^* $ (such as percolated bilayers and tubes) where the simulation cell is cubic and static --- return the behaviour of the system under a percolation enforced confinement, in the case of the continuous structures, or a kind of low density enforced confinement, as is the case with topologically closed structures. Constant pressure simulations with variable box dimensions may be employed to explore this possibility. 

\subsection{Phase diagrams close to the Janus limit}\label{res:phasediagrams}

\begin{figure}[htbp]
	\includegraphics[width=18cm]{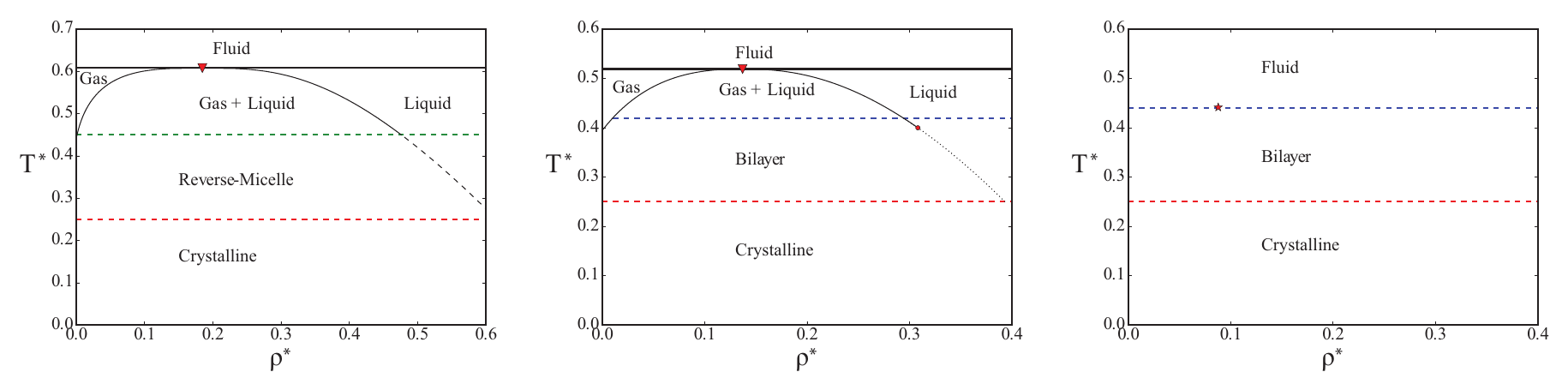} 	
	\par\bigskip
	\caption{Phase diagrams for (left to right) $ \alpha = 1.3, 1.2,$ and $ 1.1 $. As $ \sigma_h $ approaches $ \sigma_s $ the phase diagram changes to develop layered structures. This begins with the formation of reverse micelles (local segregation of $ h $ spheres) below $ T^* \approx 0.43 $ for $ \alpha = 1.3 $, followed by the formation of bilayers at a similar temperature for $ \alpha \approx 1.2 $, which increases slightly for $ \alpha = 1.1 $. The gas-liquid coexistence binodal is present for $ \alpha = 1.3 $ and $ 1.2 $ (critical point indicated by the red triangle), but is suppressed by the formation of bilayers at $ \alpha = 1.1$ (indicated by the red star). At the bottom of each plot are regions where crystalline order is observed in simulations, where additional peaks resolve in $\mathbf{g}(r)$.}
	\label{phase_diagrams}
\end{figure}

We compile the data from SUS and NVT simulations to generate a full $ T^*-\rho^* $ phase diagrams for $ \alpha = 1.3, 1.2$  and  $1.1$, that is close to the Janus limit. The diagram for $ \alpha = 1.3 $ is shown in Figure \ref{phase_diagrams} and should be contrasted with the phase diagrams appearing in Figure \ref{coex_by_alpha} where only the region close to the gas-liquid transition was depicted.

The onset of bilayer structures at essentially all densities and sufficiently low temperatures ($T^{*}\lesssim0.42$), is a clear indication of a progressive destabilisation of the gas-liquid phase separation in favour of a self-assembled bi-layered structure, in agreement with the phase diagram of the Janus limit obtained in previous
work.\cite{MUNAO2014} There the metastability of the gas-liquid transition with respect to formation of bilayer aggregates was observed via a different extrapolation method. Present results indicate this to occur in a region $1.1 \lesssim \alpha <1.2$, that is \textit{before} reaching the Janus limit. Also, given that these structures form spontaneously at low densities indicates a strong preference for any structure found across an isotherm to be dominated by the formation of bilayers and that, since the smaller $ h $ bead allows significant curvature, locally similar structures (as regards an individual particle in a bilayer) can have radically different global topologies (see, for example, snapshots in Fig \ref{large_sys_snaps}). Here an investigation of the relative stability of different topologies may be considered. Given the change in system topology on cooling, and their coincidence on the phase diagram, it would be interesting to consider the effect of field mixing on the critical behaviour.\cite{WILDING1995} A similar system with an anisotropic potential found that Ising universality is preserved despite increasing anisotropy\cite{BIANCHI2008}. However, this is not necessarily clear the case here, and warrants further investigation. 

\section{Conclusions}\label{conc}
In this study we have used state-of-the-art Monte Carlo simulations to study the phase diagram and structural properties of a system formed by size asymmetric dumbbells whose spherical components have different interaction properties. One site (denoted as $s$) is the origin of a square-well potential for all similar sites on other dumbbells, whereas the $h$ site interact with similar sites on other dumbbells via a simple hard-sphere potentials. Unlike sites (i.e. $h$ with $s$) are also considered to be simply hard-sphere interacting. Sizes of the two beads forming the dumbbell are however related via parameter $\alpha$ in such a way that a transformation between different particle descriptions can be traversed while still maintaining the system characteristic length ($ \sigma $) is constant, so that the two limits are a system of $s$ spherical particles on one end ($\alpha=2$), and a system of $h$ spherical particles on the other end ($\alpha=0$) with the rest of the space characterising the HJD.

We have focussed on the region where the $s$ bead is larger than the $h$ one (\emph{i.e.} $1< \alpha \le 2$), where the onset of a gas-liquid phase separation, favoured by the large $s$ bead, is contrasted by the tendency to self assembly, favoured by the tendency to minimise the effect of the steric hindrance of the $h$ bead and, at the same time, saturate all favorable contacts of the $s$ beads.

By starting with pure $s$ square-well fluid and gradually increasing the size of the $h$ bead, we find two distinct regimes. In the first ($1.65 < \alpha \le 2$) the gas-liquid phase diagram characteristic of a pure square-well fluid is essentially unchanged with rescaled critical temperatures and densities. Surprisingly, we find a small increase of the critical density $\rho_c^{*}$ with respect to the pure $s$ fluid, that can be explained in terms of the influence of the growing $h$ bead. At lower $\alpha$, changes in the critical temperatures and densities become more drastic, indicative of a structural change where phase separation is progressively destabilised by the formation of self-assembled bilayer structures that span essentially all densities at sufficient low temperatures. This behaviour is also observed, albeit with different aggregate structures in a structurally related anisotropic particle system \cite{AVVISATI2015a}. Our results clearly indicate that gas-liquid phase separation becomes metastable with respect to bilayer formation before actually reaching the Janus limit. 

While the present work has focussed on the $\alpha>1$ asymmetry region, it would be extremely interesting to study the $\alpha<1$ limit, that is in fact the region where experimental work has been carried out \cite{KRAFT2012}. This analysis is under way and will be reported in a future publication.

The authors would like to thank Gianmarco Muna\`{o} for discussions concerning the paper's topic and The University of Sydney for providing computational resources.



\bibliography{HJD_arxiv.bib} 

\end{document}